\documentclass[british,aip,jcp,preprint,superscriptaddress]{revtex4}
\usepackage{mathptmx}
\usepackage[T1]{fontenc}
\usepackage{color}
\usepackage{babel}
\usepackage{amstext}
\usepackage{graphicx}
\usepackage{esint}
\usepackage[unicode=true,pdfusetitle,
 bookmarks=true,bookmarksnumbered=false,bookmarksopen=false,
 breaklinks=true,pdfborder={0 0 0},backref=false,colorlinks=true]
 {hyperref}
\hypersetup{
 linkcolor=blue,citecolor=blue,filecolor=blue,urlcolor=blue}

\makeatletter
\@ifundefined{textcolor}{}
{%
 \definecolor{BLACK}{gray}{0}
 \definecolor{WHITE}{gray}{1}
 \definecolor{RED}{rgb}{1,0,0}
 \definecolor{GREEN}{rgb}{0,1,0}
 \definecolor{BLUE}{rgb}{0,0,1}
 \definecolor{CYAN}{cmyk}{1,0,0,0}
 \definecolor{MAGENTA}{cmyk}{0,1,0,0}
 \definecolor{YELLOW}{cmyk}{0,0,1,0}
}

\usepackage{doi}

\makeatother

\begin{document}

\title{Molecular weight dependent bimolecular recombination in organic solar
cells }

\author{Bronson Philippa}

\affiliation{School of Engineering and Physical Sciences, James Cook University,
Townsville 4811, Australia}

\author{Martin Stolterfoht}

\affiliation{Centre for Organic Photonics \& Electronics (COPE), School of Chemistry
and Molecular Biosciences and School of Mathematics and Physics, The
University of Queensland, Brisbane 4072, Australia}

\author{Ronald D. White}

\affiliation{School of Engineering and Physical Sciences, James Cook University,
Townsville 4811, Australia}

\author{Marrapan Velusamy}

\affiliation{Centre for Organic Photonics \& Electronics (COPE), School of Chemistry
and Molecular Biosciences and School of Mathematics and Physics, The
University of Queensland, Brisbane 4072, Australia}

\author{Paul L. Burn}

\affiliation{Centre for Organic Photonics \& Electronics (COPE), School of Chemistry
and Molecular Biosciences and School of Mathematics and Physics, The
University of Queensland, Brisbane 4072, Australia}

\author{Paul Meredith}

\affiliation{Centre for Organic Photonics \& Electronics (COPE), School of Chemistry
and Molecular Biosciences and School of Mathematics and Physics, The
University of Queensland, Brisbane 4072, Australia}

\author{Almantas Pivrikas}

\affiliation{Centre for Organic Photonics \& Electronics (COPE), School of Chemistry
and Molecular Biosciences and School of Mathematics and Physics, The
University of Queensland, Brisbane 4072, Australia}
\begin{abstract}
Charge carrier recombination is studied in operational organic solar
cells made from the polymer:fullerene system PCDTBT:PC71BM (poly{[}\emph{N}-9''-heptadecanyl-2,7-carbazole-\emph{alt}-5,5-(4',7'-di-2-thienyl-2',1',3'-benzothiadiazole){]}
: {[}6,6{]}-phenyl-C$_{70}$-butyric acid methyl ester). A newly developed
technique High Intensity Resistance dependent PhotoVoltage (HI-RPV)
is presented for reliably quantifying the bimolecular recombination
coefficient independently of variations in experimental conditions,
thereby resolving key limitations of previous experimental approaches.
Experiments are performed on solar cells of varying thicknesses and
varying polymeric molecular weights. It is shown that solar cells
made from low molecular weight PCDTBT exhibit Langevin recombination,
whereas suppressed (non-Langevin) recombination is found in solar
cells made with high molecular weight PCDTBT.
\end{abstract}
\maketitle

\section{Introduction}

Bimolecular recombination is one of the key loss mechanisms in organic
bulk heterojunction solar cells, especially in thicker devices or
those made from materials which do not possess a sufficiently high
carrier mobility\cite{Blom2007,Deibel2010,Pivrikas2010}. Recombination
coefficients are commonly compared with the prediction of Langevin
\cite{Pivrikas2005,Pivrikas2007,Deibel2010}, i.e. $\beta_{\text{L}}=e\left(\mu_{\text{p}}+\mu_{\text{n}}\right)/\epsilon\epsilon_{0}$,
where $e$ is the charge of an electron, $\mu_{\text{p}}$ ($\mu_{\text{n}}$)
is the mobility of holes (electrons), and $\epsilon\epsilon_{0}$
is the dielectric permittivity. A suppressed, non-Langevin recombination
coefficient (with $\beta<\beta_{L}$) has been reported in organic
photovoltaic blends that exhibit high performance \cite{Juska2006,Shuttle2008a,Koster2011,Clarke2012,Azimi2013}.
Suppressed recombination is desirable to ensure efficient charge extraction.
The reduction factor $\beta/\beta_{L}$ is a useful ``figure of merit''
for screening candidate photovoltaic blends to rapidly identify those
which are likely to be highly performing \cite{Pivrikas2005}.

A variety of techniques are available to study recombination dynamics.
Techniques that operate on fully operational devices (i.e. those without
blocking layers or other modifications\cite{Wetzelaer2013}) include
transient photovoltage (TPV) \cite{Shuttle2008b,Shuttle2008}, photogenerated
charge extraction by linearly increasing voltage (photo-CELIV) \cite{Juska2000,Mozer2005b},
and time-of-flight (TOF) \cite{Spear1969,Spear1974}.

TPV studies often show an apparent reaction order higher than the
expected value of two \cite{Kirchartz2012}. It has been suggested
that this is due to a concentration dependence in the recombination
coefficient \cite{Maurano2010}, recombination through trap states\cite{Foertig2012},
or the spatial separation of the carriers under open circuit conditions
\cite{Kirchartz2012}. The spatial separation at open-circuit conditions
can be reduced by studying the solar cell nearer to short-circuit
conditions, as in the photo-CELIV or TOF experiments.

Photo-CELIV can be used to study charge carrier mobility and also
the bimolecular recombination coefficient \cite{Adam2011,Egbe2007,Egbe2010,YilmazCanli2010,Gunes2010,Vijila2013}.
The recombination coefficient can be estimated from the maximum extraction
current in the photo-CELIV transient\cite{Nekrasas2012,Juska2011}.
However, this transient is influenced by experimental factors that
are not fully accounted for in the theory, such as the spatial distribution
of light absorption\cite{Juska2011}, the circuit resistance\cite{Neukom2011},
and the voltage slope\cite{Lorrmann2010}. Additionally, premature
escape of charge from the film\cite{Osterbacka1998} contributes to
the charge redistribution during the delay time\cite{Sliauzys2006},
which results in a false position of the extraction maximum and makes
the measurement unreliable. While some attention has been directed
to minimizing this issue\cite{Baumann2012}, a full compensation of
carrier redistribution is impossible due to Fermi level pinning, an
inhomogeneous electric field inside the film and strong diffusion
near the electrode where carriers are photogenerated.

Another well known technique to characterise recombination is high
intensity time-of-flight (TOF)\cite{Tiwari2009,Kokil2012,Clarke2012a}.
The recombination coefficient can be estimated from the amount of
charge extracted during a TOF experiment\cite{Pivrikas2005,Pivrikas2005a,Vijila2007}.
However, the external circuit resistance influences the extracted
charge\cite{Clarke2012}, making the measurement unreliable due to
its dependence on the experimental conditions. Previous works have
neglected the impact of the $RC$ circuit \cite{Pivrikas2005}. Here,
we resolve this issue by extending the previous work to achieve more
reliable experimental results.

In this article, we study recombination in the benchmark organic photovoltaic
system PCDTBT:PC71BM (see the experimental section for details of the
materials). We quantify the recombination in this system, and compare solar
cells made with low molecular weight PCDTBT to those made with high molecular
weight PCDTBT. Our recombination study will be conducted using a variant of
time-of-flight that we call High Intensity Resistance dependent PhotoVoltage
(HI-RPV). An exact analytic solution of the relevant differential equations is
not known, so we apply numerical simulations to show the applicability of the
technique to a variety of experimental conditions
\cite{Koster2005,Hwang2009,Neukom2011,Neukom2012,MacKenzie2012,Hanfland2013}.
The details of our numerical solver are presented in Appendix A. After
demonstrating the generality of the technique, we go on to apply it to
operational bulk heterojunction solar cells.

\section{Experimental setup}

\begin{figure}
\begin{centering}
\includegraphics[width=1\columnwidth]{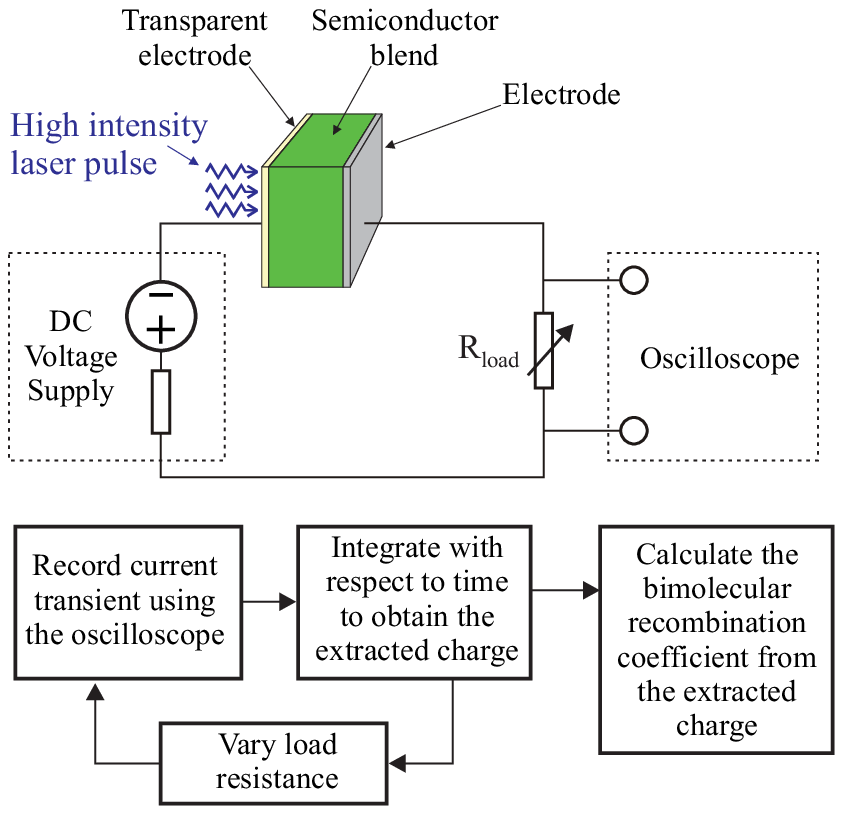}
\par\end{centering}

\protect\caption{\label{fig:Circuit-Schematic}Circuit schematic for the High Intensity
Resistance dependent PhotoVoltage (HI-RPV) experiment. Current transients
are recorded across a range of load resistances, and then integrated
to obtain the extracted charge, $Q_{\text{e}}$. The variation in
the extracted charge with resistance is used to quantify the recombination
processes and determine the bimolecular recombination coefficient.
If the device under test is an operational solar cell, then the DC
voltage supply is optional and the experiment can be done under the
solar cell's built-in field.}
\end{figure}

The experimental setup is shown in Figure \ref{fig:Circuit-Schematic}.
Similarly with time-of-flight, charges are photogenerated using a
high intensity laser, and the voltage across the load resistor is
measured with an oscilloscope. However, in contrast with traditional
time-of-flight, the measurement is repeated many times across a wide
range of load resistances. Furthermore, volume photogeneration is
desirable, and consequently operational thin-film solar cells can
be studied.

The experiment begins with the photogeneration of a large quantity
($\gg CU$) of charge carriers using an intense laser pulse. These
carriers induce a photocurrent that charges the electrodes, which
act as capacitive plates. The electrodes rapidly acquire a charge
of $CU$, where $C$ is the capacitance and $U$ is the solar cell's
built in field (or the applied voltage). Next, two processes occur
simultaneously. The first is the recombination of the photogenerated
charges, and the second is the discharge of the capacitor through
the external $RC$ circuit. If the $RC$ time is large, then the photocarriers
will completely recombine before the capacitor can discharge. Regardless of
the nature of the recombination, one can always find a resistance $R$ large enough
that the $RC$ time greatly exceeds the lifetime of charge carriers. Consequently,
in the limit of large $R$, the extracted charge will be limited to $CU$. Conversely, if
the $RC$ time is small, then the capacitor will discharge before
the carriers completely recombine, more photocurrent will flow, and
the extracted charge will exceed $CU$. In the intermediate regime,
there is an interplay between the bimolecular lifetime and the $RC$
time. We exploit this relationship in order to quantify the carrier
recombination.

\section{Device Thickness and Light Absorption Profile}

\begin{figure}
\begin{centering}
\includegraphics[width=1\columnwidth]{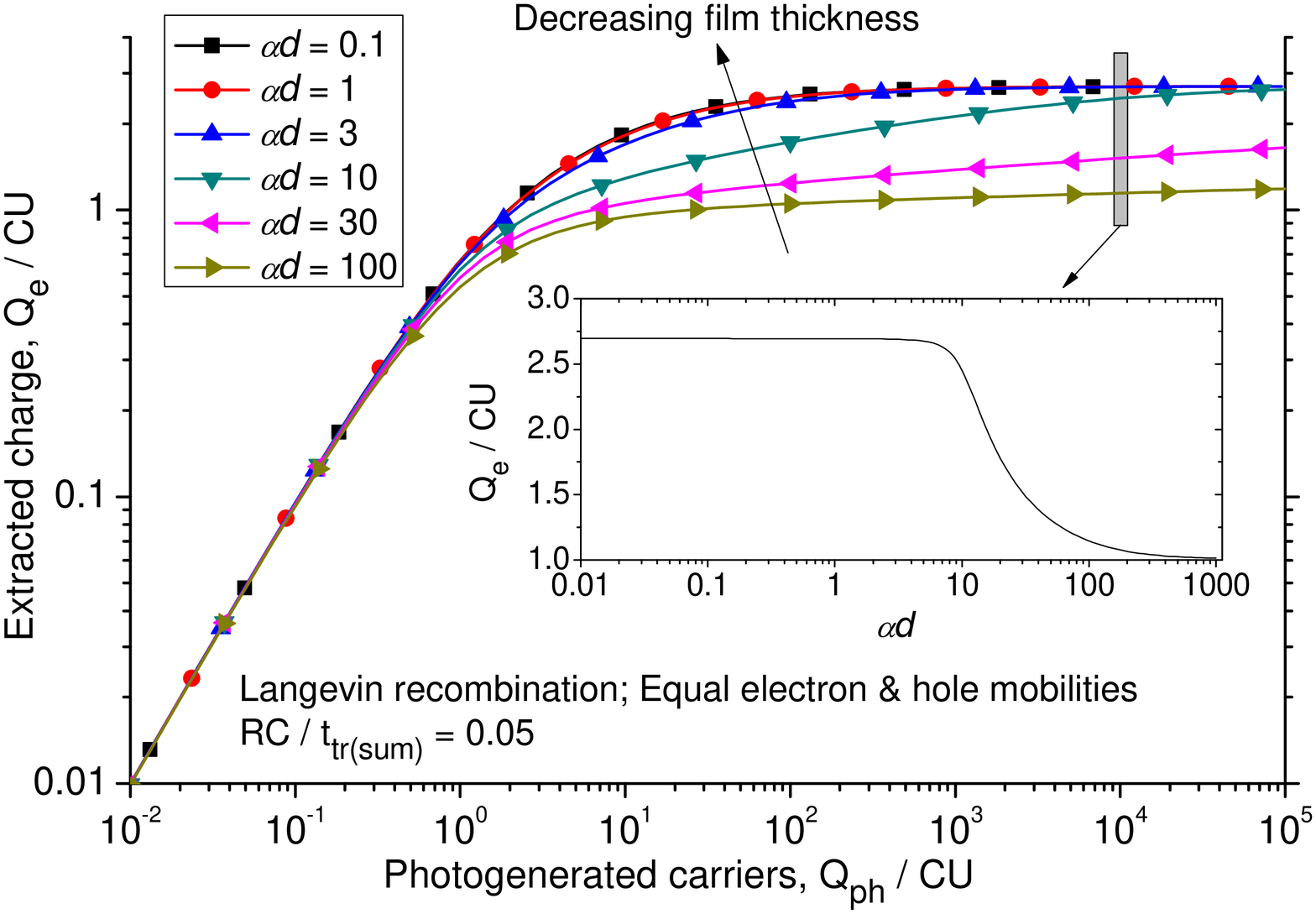}
\par\end{centering}

\protect\caption{\label{fig:Qe_vs_L_varying_alpha}The impact of the film thickness
and light absorption profile on the extracted charge. The film thickness
is incorporated within the absorption-thickness product $\alpha d$
(where $\alpha$ is the absorption coefficient and $d$ the thickness).
The inset shows the $\alpha d$ dependence in the region indicated
by the thin grey box ($Q_{\text{ph}}/CU=10^{4}$), and demonstrates
that the extracted charge is independent of the initial carrier distribution
for thin films ($\alpha d<1$). The extracted charge readily saturates
with high light intensity. This graph shows that a general theory
for thin film devices can be developed, without detailed optical modelling,
and without regard for the precise quantity of photogenerated carriers
in the saturation regime.}
\end{figure}

The simulated impact of light intensity and the optical absorption (or
photogeneration) profile is shown in Figure \ref{fig:Qe_vs_L_varying_alpha}.
We applied the Beer-Lambert law to represent the photogeneration profile,
\begin{equation}
n_{0}=p_{0}=L\alpha e^{-\alpha x},\label{eq:Beer-Lambert}
\end{equation}
where $n_{0}$ ($p_{0}$) is the initial concentration of electrons
(holes), $L$ is the light intensity in photons per unit area, $\alpha$
is the absorption coefficient at the laser wavelength, and $x$ is
the spatial coordinate. The other simulation settings are given in Appendix B.

Figure \ref{fig:Qe_vs_L_varying_alpha} shows that
the extracted charge $Q_{\text{e}}/CU$ becomes essentially independent
of $\alpha d$ when $\alpha d$ is less than 1, where $d$ is the device thickness. The inset of Figure
\ref{fig:Qe_vs_L_varying_alpha} shows the $\alpha d$ dependence
at high light intensity, demonstrating that $Q_{\text{e}}/CU$ is
essentially insensitive to the initial carrier spatial distribution,
in the case of volume generation. For example, in the case of $\alpha d=3$
(see Figure \ref{fig:Qe_vs_L_varying_alpha}), the light intensity
at the back of the device is approximately 5\% of the light intensity
at the front of the device. Such a strong inhomogeneity in the spatial
distribution does not meaningfully affect the extracted charge.

Physically, the insignificance of the initial spatial distribution
is caused by bimolecular recombination. The bimolecular recombination
process will be more rapid in regions of higher light intensity, and
slower in regions of lower light intensity. This will, in effect,
``smooth'' the carrier distribution across the device, erasing the
initial spatial distribution. More precisely, the carrier concentration
at early times is given by $n(t)=\left(n_{0}^{-1}+\beta t\right)^{-1}$,
where $n_{0}$ is the initial carrier concentration \cite{Lampert1970}.
In the limit of very large $n_{0}$, the dependence on the initial
condition vanishes {[}$n(t)\approx\left(\beta t\right)^{-1}${]}.
This explains why the absorption profile is irrelevant at high light
intensities.

In summary, to first order, detailed optical modelling to account
for exact carrier distribution in operational solar cells is not necessary,
since the precise spatial distribution of carriers is rapidly erased
by bimolecular recombination. Therefore, the HI-RPV technique can
be applied to thin film devices ($\alpha d\leq1$) without concern
for optical interference.

Since the technique is insensitive to the light absorption profile,
we will remove $\alpha d$ from the set of parameters being tested,
and approximate the initial condition by perfectly uniform carrier
generation. All subsequent numerical calculations are performed with
this simplified uniform initial condition, rather than the Beer-Lambert
law.

\section{Circuit Resistance}

\begin{figure}
\begin{centering}
\includegraphics[width=1\columnwidth]{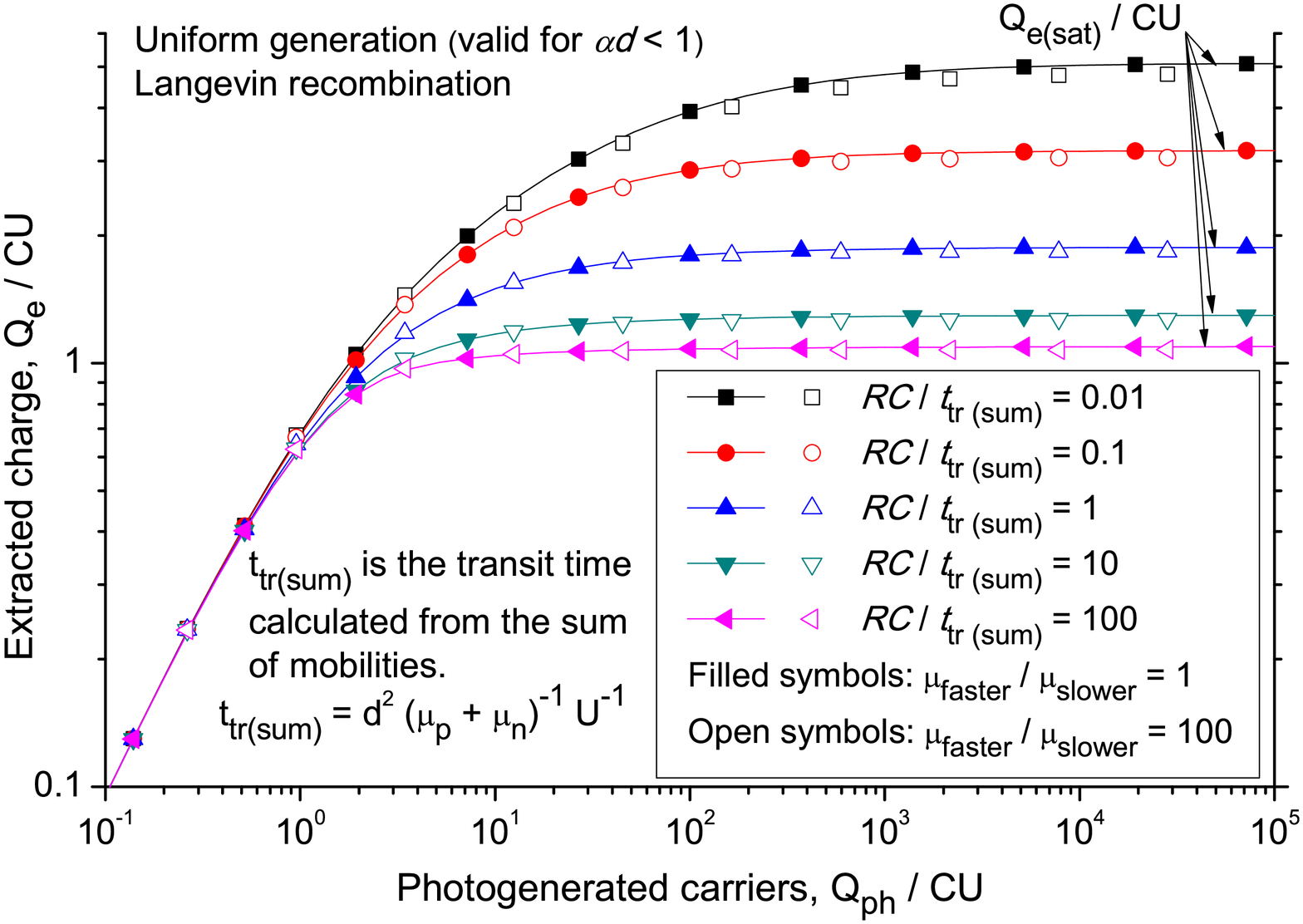}
\par\end{centering}

\protect\caption{\label{fig:Qe_vs_L_varying_R}The impact of the circuit resistance
on the extracted charge from simulated resistance dependent photovoltage
experiments. Filled symbols with lines show balanced mobilities ($\mu_{\text{faster}}/\mu_{\text{slower}}=1$);
open symbols without lines show strongly unbalanced mobilities ($\mu_{\text{faster}}/\mu_{\text{slower}}=100$).
The two are very similar, because the normalisation scale for the
circuit $RC$ time minimises the effect of the mobility ratio. The
saturation value $Q_{\text{e(sat)}}/CU$ depends almost entirely upon
the normalised resistance. These results demonstrate that the load
resistance needs to be accounted for to correctly measure the recombination
coefficient.}
\end{figure}

In a HI-RPV experiment, the circuit resistance is varied over many
orders of magnitude in order to observe the dynamical interaction
between the known circuit $RC$ time and the unknown bimolecular lifetime.

We examined the impact of the circuit resistance using our simulations,
as shown in Figure \ref{fig:Qe_vs_L_varying_R}. Importantly, we observe
that more charge can be extracted at lower resistances. A smaller
resistance allows the charge extraction to complete in a shorter time,
so that less recombination occurs, and the overall extracted charge
is higher.

The faster carrier mobility is normalised out of the simulation by
the system of units (as described in Appendix A). However, it is
necessary to specify the ratio of carrier mobilities $\mu_{\text{faster}}/\mu_{\text{slower}}$.
To confirm that variation in this ratio will not interfere with the
measurement, Figure \ref{fig:Qe_vs_L_varying_R} shows the case of
balanced mobilities ($\mu_{\text{faster}}/\mu_{\text{slower}}=1$)
with filled symbols and lines and strongly unbalanced mobilities ($\mu_{\text{faster}}/\mu_{\text{slower}}=100$)
with open symbols and no lines. This covers a wide range of mobility
ratios to examine the variation that might be expected to occur in
practice. The two cases (balanced mobilities and strongly unbalanced
mobilities) are essentially indistinguishable, as shown in Figure
\ref{fig:Qe_vs_L_varying_R}. We explain this insensitivity as follows.
The amount of extracted charge $Q_{\text{e}}$ is primarily controlled
by the recombination. The Langevin recombination rate is proportional
to the\emph{ sum} of carrier mobilities. The relevant time scale for
this process is $t_{\text{tr(sum)}}\equiv d^{2}/\left(\mu_{\text{p}}+\mu_{\text{n}}\right)U$.

Figure \ref{fig:Qe_vs_L_varying_R} shows that the extracted charge
saturates at high light intensities to a value that we call $Q_{e(\text{sat})}/CU$,
as indicated by the arrows. Therefore, if the HI-RPV experiment is
operated in this saturation regime, the amount of extracted charge
does not depend on the laser power which is applied. The extracted
charge is also independent of the carrier mobility ratio (Figure \ref{fig:Qe_vs_L_varying_R})
and the light absorption profile (Figure \ref{fig:Qe_vs_L_varying_alpha}).
Consequently, the only parameters remaining to be quantified are the
circuit resistance and the bimolecular recombination coefficient.

\begin{figure}
\begin{centering}
\includegraphics[width=1\columnwidth]{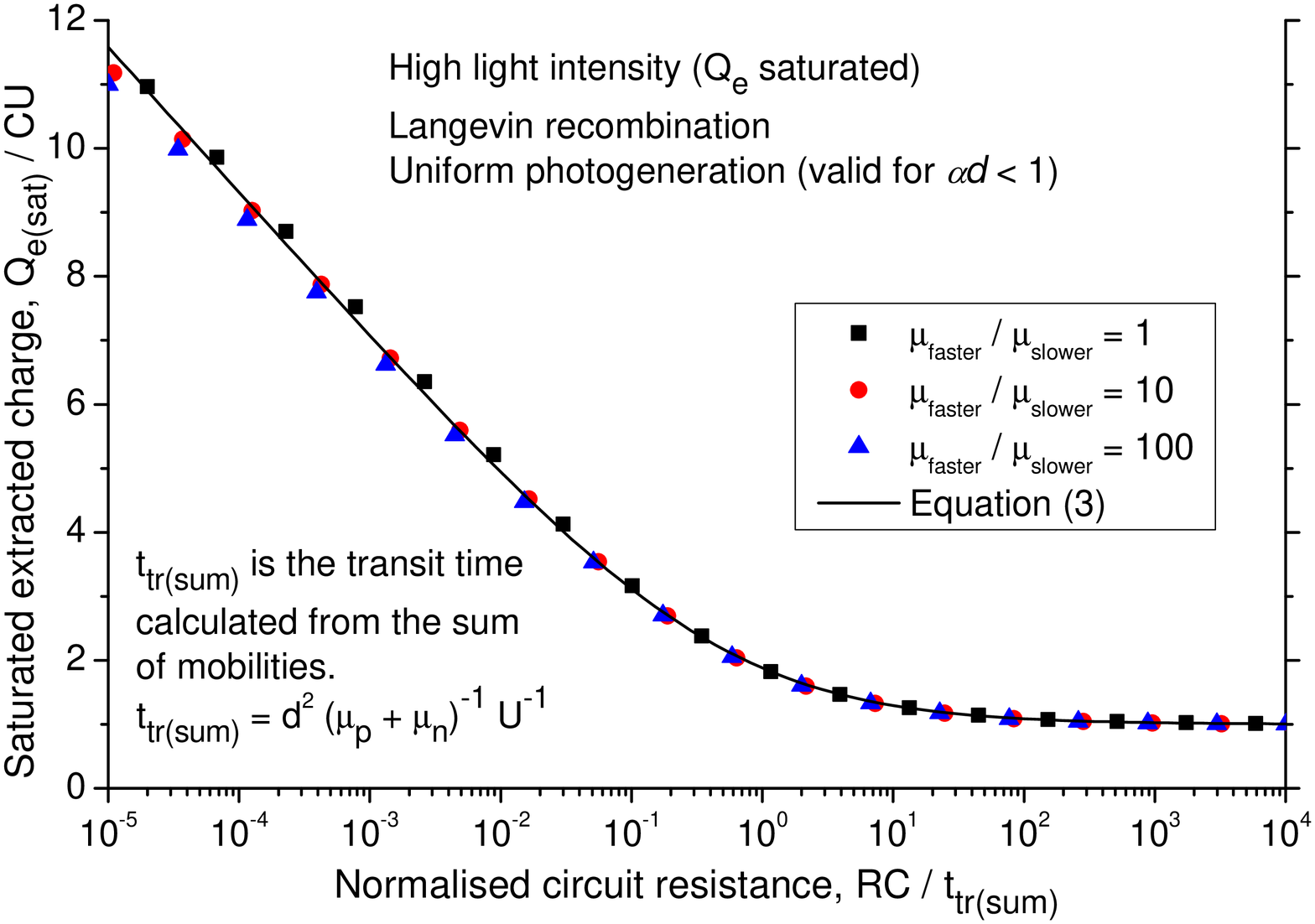}
\par\end{centering}

\protect\caption{\label{fig:QeSat(R)_sims}Simulations of the impact of load resistance
on the extracted charge from thin film devices with Langevin recombination
at varying mobility ratios. Points are calculated from simulations
at high light intensity ($Q_{\text{ph}}/CU=10^{6}$, although the
precise value is unimportant because of the saturation in the extracted
charge $Q_{\text{e}}$, as shown in Figure \ref{fig:Qe_vs_L_varying_R}).
The ratio of carrier mobilities does not affect the extracted charge,
so HI-RPV measurements can be applied equally to systems with balanced
mobilities and systems with strongly unbalanced mobilities. }
\end{figure}

The impact of the circuit resistance is shown in Figure \ref{fig:QeSat(R)_sims}.
If the normalised resistance is small, the extracted charge $Q_{\text{e(sat)}}$
can exceed the charge on the electrodes $CU$ by an order of magnitude
or more, even in the presence of Langevin recombination. The TOF experiment
under these circumstances is therefore misleading, especially if comparing
two systems with different values of the normalised resistance, $RC/t_{\text{tr(sum)}}$.
We resolve this problem by introducing the HI-RPV technique. We firstly
develop predictions for Langevin systems, and then in the following
section extend this to the general case.

Figure \ref{fig:QeSat(R)_sims} shows a single universal curve that
all Langevin systems should obey. We developed an empirical equation
to describe this curve by arbitrarily choosing an appropriate functional
form that would give a logarithmic dependence at small $R$ (as shown
in Figure \ref{fig:QeSat(R)_sims}), and would saturate to 1 at large
$R$ (as also shown in Figure \ref{fig:QeSat(R)_sims}),
\begin{equation}
\frac{Q_{\text{e(sat)}}}{CU}=1+p_{1}\log\left[1+p_{2}\left(\frac{t_{\text{tr(sum)}}}{RC}\right)^{p_{3}}\right].\label{eq:QeSat(R)_general}
\end{equation}
We used non-linear least squares regression to calculate the coefficients
$p_{i}$ from the simulation results in Figure \ref{fig:QeSat(R)_sims}.
The result is:

\begin{equation}
\frac{Q_{e\text{(sat)}}}{CU}=1+1.8\log\left[1+0.63\left(\frac{t_{\text{tr(sum)}}}{RC}\right)^{0.55}\right],\label{eq:QeSat(R)_Langevin}
\end{equation}
which is valid for Langevin recombination and thin films. Equation
(\ref{eq:QeSat(R)_Langevin}) is plotted against the simulation results
in Figure \ref{fig:QeSat(R)_sims}, demonstrating excellent agreement.

The purpose of Eq. (\ref{eq:QeSat(R)_Langevin}) is to determine the
type of recombination present in a thin film device; for example,
one could plot this equation alongside measured data in order to determine
whether the recombination is of the Langevin type. This is important,
since recombination orders higher than two have been experimentally
observed \cite{Kirchartz2012}, and it is necessary to identify the
type of recombination dynamics that might apply to the system being
studied. A plot of extracted charge versus resistance (Figure \ref{fig:QeSat(R)_sims})
will follow the form of Eq. (\ref{eq:QeSat(R)_Langevin}) if Langevin
recombination is dominant. In contrast, if there is a higher order
of recombination, then the carrier concentration will decay according
to a different time dependence, and the functional form of the extracted
charge versus resistance will change. If the recombination is stronger
than Langevin, the experimental data will lie \emph{below} the line.
On the other hand, if the dominant form of recombination is slower
than Langevin, then less recombination will occur and the experimental
data will lie \emph{above} the line.

We will show below that our experimental data can be described by
a bimolecular recombination process with a Langevin reduction prefactor.
We do not exclude the possibility of higher-order effects such as
a concentration-dependent recombination coefficient \cite{Maurano2010},
but these are not necessary to explain our data. Therefore, in the
following section, we extend our theory to systems with suppressed
(non-Langevin) recombination of purely second order.

\section{Bimolecular Recombination Coefficient}

\begin{figure}
\centering{}\includegraphics[width=1\columnwidth]{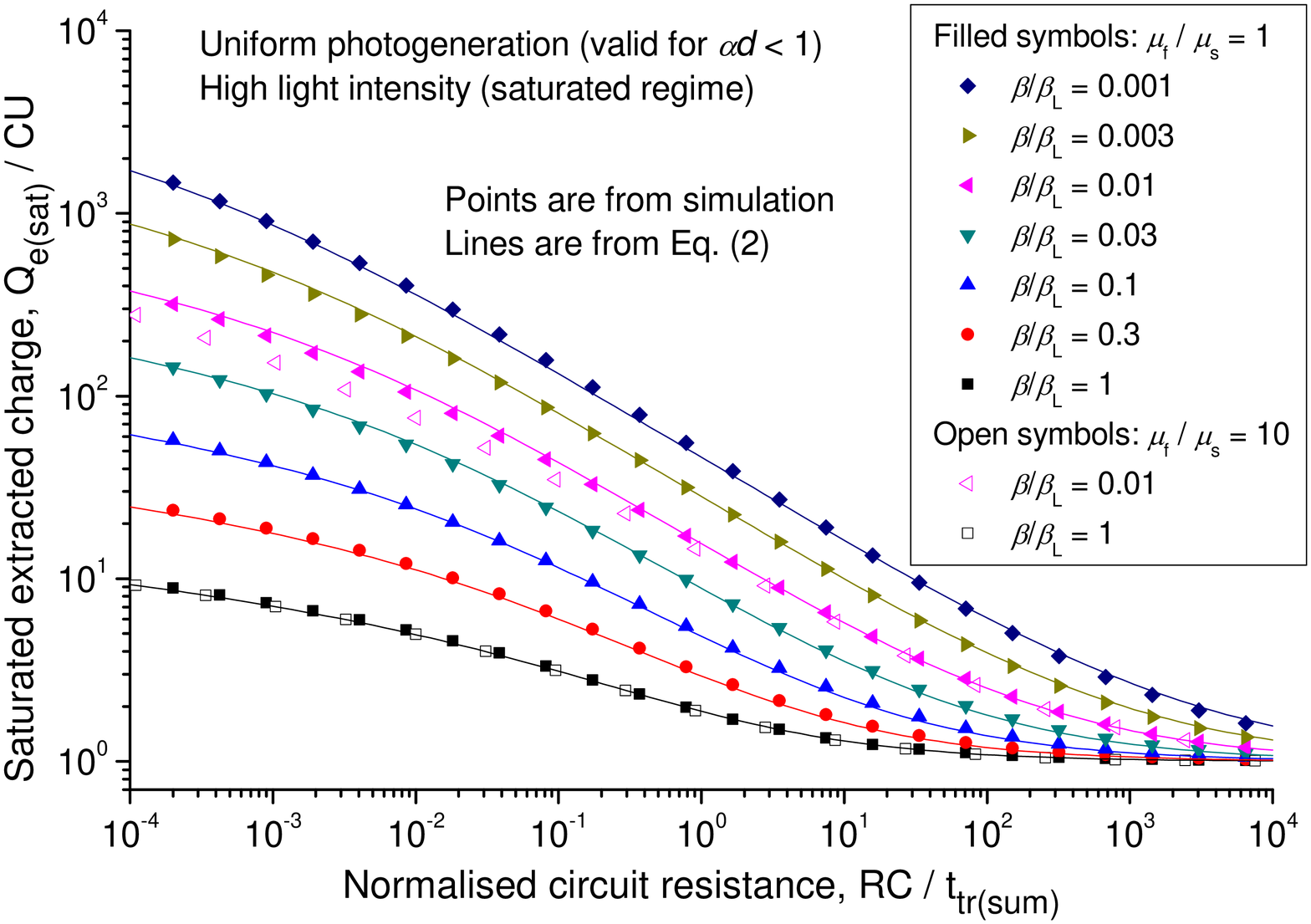}\protect\caption{\label{fig:QeSat(R,beta)_sims}Numerically predicted extracted charge
as a function of load resistance in high light intensity resistance
dependent photovoltage (HI-RPV) experiments for different recombination
coefficients $\beta/\beta_{\text{L}}$. The extracted charge shown
in this figure is calculated at the highest light intensities where
the extracted charge saturates, as shown in Fig. \ref{fig:Qe_vs_L_varying_alpha}.
The points are from simulations, whereas the lines are Eq. (\ref{eq:QeSat(R)_general})
evaluated for each respective value of $\beta/\beta_{\text{L}}$.
This graph presents numerical predictions to be used when measuring
the recombination coefficient $\beta/\beta_{L}$ experimentally from
HI-RPV in systems without deep traps.}
\end{figure}

In order to develop a tool for convenient experimental quantification
of the recombination coefficient ($\beta/\beta_{L}$), we applied
numerical simulations to predict the amount of extracted charge as
a function of $\beta/\beta_{L}$. These simulations are plotted in
Figure \ref{fig:QeSat(R,beta)_sims}. As expected, the amount of extracted
charge increases dramatically in the presence of non-Langevin recombination.
To confirm that our technique remains valid, we checked that non-Langevin
devices also exhibit saturation at high light intensity, and that
the extracted charge is independent of the optical absorption profile
for thin films ($\alpha d<1$). We found that systems with strongly
suppressed recombination ($\beta\ll\beta_{\text{L}}$) exhibit a stronger
dependence on the mobility ratio than Langevin systems. The more unbalanced
the mobilities, the less charge can be extracted. A representative
example ($\mu_{\text{faster}}/\mu_{\text{slower}}=10$) is plotted
in Figure \ref{fig:QeSat(R,beta)_sims} with open symbols.

We are now ready to specify how the HI-RPV technique can be applied.
The recombination coefficient can be determined by comparing measurements
of the extracted charge against the simulation results in Figure \ref{fig:QeSat(R,beta)_sims}.
This approach is valid for any thin film ($\alpha d<1$) device. Importantly,
this technique is not hindered by the RC-dependence that affects traditional
high intensity TOF \cite{Pivrikas2005,Clarke2012}, because the impact
of the $RC$ time constant on the extracted charge is accounted for
on the horizontal axis of Figure \ref{fig:QeSat(R,beta)_sims}. However,
for accurate measurements, it is necessary to reach the regime where
$RC/t_{\text{tr(sum)}}\ll1$. This may not be possible in extremely
high mobility materials, especially when the series resistances are
included in $R$. Ideally, $R$ should be varied over many orders
of magnitude.

As an alternative to visual inspection of the graph, we can also specify
an empirical equation that describes the data in Figure \ref{fig:QeSat(R,beta)_sims}.
We started with the general functional form {[}Eq. (\ref{eq:QeSat(R)_general}){]}
and applied a procedure similar to that described earlier for the
Langevin case. With least squares regression,
we found the parameters $p_{i}$ as a function of $\beta/\beta_{\text{L}}$.
Finally, we parametrised the $p_{i}$ values as follows, choosing
an arbitrary functional form that best described the data:
\begin{eqnarray}
p_{1} & = & 1.829\left(\frac{\beta}{\beta_{\text{L}}}+0.0159\sqrt{\frac{\beta}{\beta_{\text{L}}}}\right)^{-1}\label{eq:p1}\\
p_{2} & = & 0.63\left(\frac{\beta}{\beta_{\text{L}}}\right)^{0.407}\label{eq:p2}\\
p_{3} & = & 0.55\left(\frac{\beta}{\beta_{\text{L}}}\right)^{0.0203}.\label{eq:p3}
\end{eqnarray}
These functional forms were found to obtain the best fit to the simulated
results.

Figure \ref{fig:QeSat(R,beta)_sims} shows the simulation results
compared with Eq. (\ref{eq:QeSat(R)_general}) with the parameters
(\ref{eq:p1})-(\ref{eq:p3}). A good agreement is demonstrated for
balanced mobilities; if the mobilities are unbalanced then Eq. (\ref{eq:QeSat(R)_general})
will slightly overestimate the extracted charge.

These equations are a convenient tool to analyse experimental data.
For example, to determine the recombination coefficients for the data
presented below, we set up a spreadsheet table to compare the model
with experimental data and thereby estimate the bimolecular recombination
coefficient.

In order to confirm the validity of the newly presented HI-RPV technique,
we have compared its results in various systems with other techniques
including photo-CELIV, double injection transients, plasma extraction,
and steady-state IVs. The results are in agreement, given the limitations
of each technique. These limitations must be carefully considered
when comparing measurements, which is why we have developed the present
HI-RPV approach.

\section{Experimental Measurements}

\begin{figure}
\begin{centering}
\includegraphics[width=1\columnwidth]{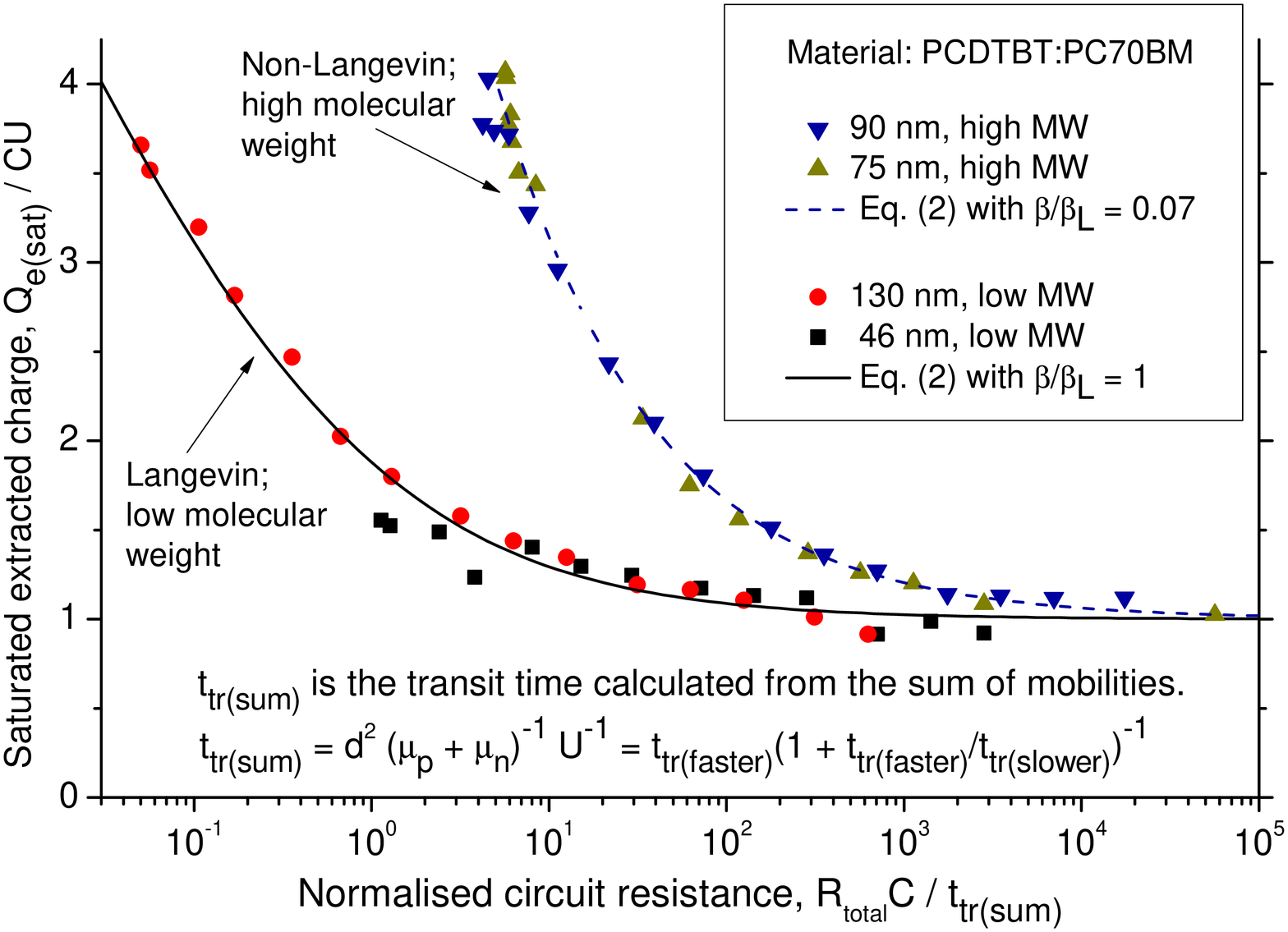}
\par\end{centering}

\protect\caption{\label{fig:QeSat(R)_experimental}Experimentally measured extracted
charge as a function of circuit resistance obtained using the HI-RPV
technique. Films made with the low molecular weight polymer exhibit
Langevin recombination, whereas films containing the high molecular
weight polymer exhibit suppressed non-Langevin recombination. Non-Langevin
recombination is beneficial to solar cell performance, indicating
the importance of material quality in device fabrication.}
\end{figure}

We manufactured bulk heterojunction solar cells with the donor:acceptor
blend poly{[}\emph{N}-9''-heptadecanyl-2,7-carbazole-\emph{alt}-5,5-(4',7'-di-2-thienyl-2',1',3'-benzothiadiazole){]}
(PCDTBT) and {[}6,6{]}-phenyl-C$_{71}$-butyric acid methyl ester
(PC70BM). This blend, PCDTBT:PC70BM, has previously been reported
to exhibit near to Langevin recombination\cite{Clarke2012a}. Two
sources of PCDTBT were used. A low molecular weight batch ($\overline{\text{M}}_{\text{n}}$
= 4.3 kDa, $\overline{\text{M}}_{\text{w}}$ = 12.1 kDa, PDI = 2.8,
obtained in 1,2,4-trichlorobenzene at 140 $^{\circ}$C) was synthesized
in our laboratory following the Suzuki cross-coupling protocols previously
described\cite{Blouin2008}. A high molecular weight batch ($\overline{\text{M}}_{\text{n}}$
= 22.7 kDa, $\overline{\text{M}}_{\text{w}}$ = 122.2 kDa, PDI = 5.4)
was purchased from the SJPC Group.

The fabrication of the solar cells followed a previously described
procedure\cite{Armin2012a}. 15 $\Omega$/sq. Indium Tin Oxide (ITO)
coated glass substrates patterned by photolithography (Kintec) were
cleaned by sonicating in sequence with alconox (detergent), de-ionised
water, acetone, and\emph{ iso-}propanol for 10 minutes. The cleaned
substrates were coated with a 20 nm layer of poly(3,4-ethylenedioxythiophene):poly(styrene
sulfonate) (PEDOT:PSS) by spin casting at 5000 rpm for 60 sec. The
PEDOT:PSS layer was baked for 10 minutes at 170 $^{\circ}$C. A solution
of PCDTBT and commercially purchased PC70BM (Nano-C) with a mass ratio
of 1:4 was prepared at a total concentration of 20 mg/mL in anhydrous
1,2-dichlorobenzene. This solution was deposited by spin coating on
top of the PEDOT:PSS layer after filtration. Two substrates were prepared
from the low molecular weight batch with active layer thicknesses
of 46 nm and 130 nm, respectively. From the high molecular weight
batch, two additional substrates were made with active layer thicknesses
of 75 nm and 90 nm. Thicknesses were measured by a Veeco Dektak150
profilometer. Slow drying was performed after spin coating by placing
the coated film in a partially opened petri dish for 2 hours. Finally,
a 100 nm aluminium layer was deposited by thermal evaporation under
a $10^{-6}$ mbar vacuum. The device areas were 0.035 $\text{cm}^{2}$
with three devices per substrate. The low molecular weight material
produces solar cells with power conversion efficiencies (PCE) of approximately
4\%; whereas optimised solar cells made from the high molecular weight
material have PCEs in excess of 6\%\cite{Pandey2012}. Transit times
were measured using low light intensity resistance dependent photovoltage
\cite{Philippa2013}; the mobilities were $\mu_{\text{low MW}}\approx8\times10^{-5}\,\text{cm}^{2}\text{V}^{-1}\text{s}^{-1}$
and $\mu_{\text{high MW}}\approx2\times10^{-3}\,\text{cm}^{2}\text{V}^{-1}\text{s}^{-1}$,
demonstrating greatly improved charge transport in the latter devices. Further work
would be needed to identify the underlying mechanism for this change. We note that a
strong dependence of mobility on molecular weight has been observed in other polymers
in the past \cite{Kline2003}.

HI-RPV measurements were performed using a pulsed third-harmonic Nd:YAG
laser (Quantel Brio) working at a wavelength of 355 nm and pulse duration
of 5 ns. At 355 nm, the absorption coefficient of this blend\cite{Park2009}
is $8\times10^{4}\ \text{cm}^{-1}$, which gives $\alpha d$ values
of 0.37 for the thinnest device (46 nm) and 1.0 for the thickest device
(130 nm). The laser beam was attenuated using a neutral density filter
set. No external voltage was applied; instead, the transients were
driven by the solar cells' built-in field. The signal was recorded
by a digital storage oscilloscope (LeCroy Waverunner A6200).

We performed HI-RPV with load resistances in the range from 1 $\text{\ensuremath{\Omega}}$
to 1 M$\text{\ensuremath{\Omega}}$. The results are plotted in Figure
\ref{fig:QeSat(R)_experimental}. This graph demonstrates the application
of the HI-RPV technique. It is important to note that the resistance
value $R$ on the horizontal axis is the \emph{complete} circuit resistance,
calculated as the sum of the load resistance and the solar cell series
resistance. The experimental data is plotted together with the predicted
curve from Eq. (\ref{eq:QeSat(R)_general}) with parameters (\ref{eq:p1})-(\ref{eq:p3}).
The measured extracted charge behaves as expected and as predicted
by the simulations. The extracted charge decreases with increasing
resistance until it saturates to $Q_{\text{e(sat)}}/CU=1$. To determine
the recombination strength, the coefficient $\beta/\beta_{L}$ was
adjusted until the predicted curves matched the experimental data.

Our results indicate that low molecular weight devices exhibit Langevin-type
recombination, while the high molecular weight devices exhibit non-Langevin
recombination with $\beta/\beta_{L}\approx0.07$. Photo-CELIV measurements
applied to the same devices demonstrated Langevin and non-Langevin
recombination, respectively, supporting our results. However, photo-CELIV
is subject to various limitations, as we discussed in the Introduction,
and so we developed HI-RPV for the detailed study. The strong change
in the recombination strength likely contributes to the improved power
conversion efficiency of the high molecular weight blend. It has previously
been reported that PCDTBT solar cell performance improves with increasing
molecular weight\cite{Wakim2009}. Our results indicate that suppressed
recombination may be the mechanism behind this performance trend,
and hence the molecular weight is a parameter that should be considered
when optimising solar cell performance. There may be further performance
improvements to be gained by identifying the molecular weight at which
the recombination is minimised.

A previous study of recombination in PCDTBT solar cells\cite{Clarke2012a}
reported reduction factors in the range of $\beta/\beta_{L}=0.3$
to $\beta/\beta_{L}=1$ depending upon the device thickness. Thinner
devices were reported to exhibit more strongly reduced recombination.
Thickness dependencies cannot be reliably studied using time-of-flight
because variations in the thickness influence parameters such as the
device capacitance, the $RC$ time, the transit time, the optical
absorption profile, and the amount of extracted charge. Consequently,
with time-of-flight it is difficult to eliminate the dependence on
the experimental parameters. In contrast, HI-RPV accounts for these
effects. We did not observe any thickness dependence, although the
range of thicknesses measured here is less than that in the previous
study\cite{Clarke2012a}.

Further work is necessary in order to clarify the origin of this molecular
weight dependence, as well as any dependence on other parameters such
as polydispersity, impurity density, and conjugation length. The novel
HI-RPV technique will be beneficial for such future work.

\section{Conclusion}

We studied recombination in the organic photovoltaic system PCDTBT:PC70BM,
and observed that devices made with a higher molecular weight polymer
exhibit suppressed recombination relative to devices made with a lower
molecular weight polymer. Our results highlight the importance of
material quality for fabrication of high efficiency organic solar
cells. We developed and implemented a theoretical framework for the
novel High Intensity Resistance dependent PhotoVoltage (HI-RPV) technique,
which allows recombination measurements that are independent of the
experimental conditions, resolving a key weakness of previous time-of-flight
based techniques. A key advantage of HI-RPV is its independence on
the light absorption profile in thin films, making it applicable to
operational devices.

\section*{Acknowledgements}

We thank Anton Bavdek for helpful assistance with the experimental
work. We thank the James Cook University High Performance Computing
Centre for computational resources. A.P. is the recipient of an Australian
Research Council Discovery Early Career Researcher Award (Projects:
ARC DECRA DE120102271, UQ ECR59-2011002311 and UQ NSRSF-2011002734).
P.M. and P.L.B. are UQ Vice Chancellor's Senior Research Fellows.
B.P. is funded by an Australian Postgraduate Award (APA). M.S. is
funded by a University of Queensland International scholarship (UQI).
We acknowledge funding from the University of Queensland (Strategic
Initiative - Centre for Organic Photonics \& Electronics), the Queensland
Government (National and International Research Alliances Program)
and the ARC Centre of Excellence for Antimatter-Matter Studies. This
work was performed in part at the Queensland node of the Australian
National Fabrication Facility (ANFF-Q) - a company established under
the National Collaborative Research Infrastructure Strategy to provide
nano and microfabrication facilities for Australia's researchers.

\appendix

\section{Numerical Drift-Diffusion Solver}

Our simulations take an effective medium approach to model device-scale
behaviour, an approach which is commonly used for organic solar cell
simulation \cite{Koster2005,Hwang2009,Neukom2011,Neukom2012,MacKenzie2012,Hanfland2013}.
We consider the situation where the films are not doped and there
is no film charging due to deep traps whose release times are longer
than the transit time. These assumptions are typically met in high
efficiency devices.

We apply one dimensional continuity equations for electron and hole number
densities\cite{Koster2005,Hwang2008,Rogel-Salazar2009,Hwang2009,Neukom2011}.
These are coupled to the Poisson equation, to incorporate the effects
of space charge. All quantities are scaled such that they are dimensionless.
We denote dimensionless quantities with a prime. The non-dimensionalisation
is similar to that used by Ju\v{s}ka\emph{ et al}\cite{Juska1994,Juska1995}.

The length scale is the film thickness: $x'\equiv x/d$. The time
scale is the transit time calculated for the fastest mobility: $t'\equiv t/t_{\text{tr}}$.
The voltage scale is the applied voltage: $U'\equiv U/U_{\text{applied}}$.
This system of units requires that the normalised faster carrier mobility
is $\mu_{\text{faster}}'=1$.

The charge scale is the charge on the electrodes: $Q'\equiv Q/CU$.
The number density scale is $CU$ per volume: $n'\equiv enSd/CU$,
where $S$ is the surface area of the device. The current scale is
$CU$ per transit time: $j'\equiv jt_{tr}/CU$. The circuit resistance
is expressed internally in the simulations by $R'\equiv RC/t_{\text{tr}}$,
but everywhere in this article, we present it instead as $R'\equiv RC/t_{\text{tr(sum)}}$.
This is because the scaling with respect to the\emph{ sum} of mobilities
eliminates most of the mobility ratio dependence in $Q_{\text{e}}$,
as shown in Figure \ref{fig:Qe_vs_L_varying_R}.

The Einstein relation for diffusion gives a dimensionless temperature
$T'=kT/eU_{\text{applied}}$. The recombination coefficient is normalised
to the Langevin rate: $\beta'\equiv\beta/\beta_{L}$.

The model equations for the semiconductor bulk are:
\begin{eqnarray}
j_{\text{p}}' & = & \mu_{\text{p}}'E'p'-\mu_{\text{p}}'T'\frac{\partial p'}{\partial x'}\\
j_{\text{n}}' & = & \mu_{\text{n}}'E'n'+\mu_{\text{n}}'T'\frac{\partial n'}{\partial x'}
\end{eqnarray}
\begin{eqnarray}
\frac{\partial p'}{\partial t'}+\frac{\partial j_{\text{p}}'}{\partial x'} & = & -\beta'\left(\mu_{\text{p}}'+\mu_{\text{n}}'\right)n'p'\\
\frac{\partial n'}{\partial t'}-\frac{\partial j_{\text{n}}'}{\partial x'} & = & -\beta'\left(\mu_{\text{p}}'+\mu_{\text{n}}'\right)n'p'
\end{eqnarray}
\begin{eqnarray}
\frac{\partial^{2}U'}{\partial\left(x'\right)^{2}} & = & n'-p'\\
E' & = & -\frac{\partial U'}{\partial x'}.
\end{eqnarray}

The boundary conditions for Poisson's equation are:
\begin{eqnarray}
U'(t,0) & = & V'\\
U'(t,1) & = & 0,
\end{eqnarray}
where $V'$ is the voltage across the semiconductor:
\begin{eqnarray}
\frac{dV'}{dt'} & = & \frac{1-V'}{R'}-j_{\text{c}}'\\
j_{\text{c}}' & = & \int_{0}^{1}j_{\text{p}}'(x)+j_{\text{n}}'(x)\ dx.
\end{eqnarray}

The boundary conditions for the number density are as follows. We
use a finite volume method, so the boundary conditions for the transport
equations are expressed in terms of the fluxes $j_{\text{p}}'$ and
$j_{\text{n}}'$ at each electrode. Since the RPV experiment is conducted
under reverse bias, we assume no injection is possible. This immediately
sets two such edge fluxes to zero. The other two represent charge\emph{
extraction} and are described by the local drift current $j_{\text{p}}'=\mu_{\text{p}}'E'p'$
(and similarly for electrons).

The initial condition for the number density is Eq. (\ref{eq:Beer-Lambert})
in normalised units:
\begin{equation}
n'(0,x')=p'(0,x')=L'\alpha'e^{-\alpha'x'},
\end{equation}
with $Q_{\text{ph}}'=L'\left(1-e^{-\alpha'}\right)$; or alternatively,
by the condition of uniform generation
\begin{equation}
n'(0,x')=p'(0,x')=Q_{\text{ph}}'.
\end{equation}
The initial condition for voltage is $V'=1$.

The spatial discretisation of these equations was performed using
the finite volume method. Number densities are defined at cell midpoints,
whereas the fluxes and the electric field are defined on the cell
boundaries. This results in a large system of coupled ODEs in time.
We implemented these in Matlab, and found that the\emph{ ode15s} solver
usually provides the best performance out of all the standard Matlab
ODE solvers.

\section{Simulation settings for Figure 2}

The light intensity is represented by
the quantity of photogenerated charge carriers
$Q_{\text{ph}}=L\left(1-e^{-\alpha d}\right)$, which is the integral of Eq.
(\ref{eq:Beer-Lambert}) over the device. We selected a fixed circuit
resistance, $RC/t_{\text{tr(sum)}}=0.05$, where $R$ is the resistance of the
circuit external to the device, $C$ is the device capacitance, and
$t_{\text{tr(sum)}}\equiv
d^{2}\left(\mu_{\text{p}}+\mu_{\text{n}}\right)^{-1}U^{-1}$ is an effective
transit time calculated from the sum of carrier mobilities crossing a film of
thickness $d$ under a voltage $U$. The bimolecular recombination was given by
the Langevin rate ($\beta/\beta_{L}=1$). The simulations were conducted with
equal electron and hole mobilities; however, the results are essentially
unchanged if the mobilities are not equal (as shown in Figure
\ref{fig:Qe_vs_L_varying_R}).

\bibliographystyle{apsrev4-1}

%

\end{document}